\documentclass[conference]{IEEEtran}  % Comment this line out if you need a4paper
\IEEEoverridecommandlockouts                                     % Needed to meet printer requirements.
\usepackage{cite}
\usepackage{amsmath,amssymb,amsfonts,verbatim}
\usepackage{graphicx}
\usepackage{textcomp}
\usepackage{mathrsfs}
\usepackage{epsfig}
\usepackage{epsf}
\usepackage{theorem}
\usepackage{graphics}
\usepackage[active]{srcltx}
\usepackage{tikz}
\usepackage{pifont}
\usepackage{tabulary}
\usepackage{float}
\usepackage{caption}
\usepackage{subcaption}
\usepackage{multirow}
\def\BibTeX{{\rm B\kern-.05em{\sc i\kern-.025em b}\kern-.08em
    T\kern-.1667em\lower.7ex\hbox{E}\kern-.125emX}}

\title{Real Time Video based Heart and Respiration\\ 
	Rate Monitoring\\
}

\author{\IEEEauthorblockN{Jafar Pourbemany\IEEEauthorrefmark{1}, Almabrok Essa, and Ye Zhu}
\IEEEauthorblockA{\textit{Department of Electrical Engineering and Computer Science} \\
\textit{Cleveland State University, Cleveland, OH, USA}\\
\IEEEauthorrefmark{1}pourbemany@ieee.org; a.essa@csuohio.edu; y.zhu61@csuohio.edu}
}

\begin{document}
\maketitle
\pagestyle{empty}
\thispagestyle{empty}
%%%%%%%%%%%%%%

%%%%%%%%%%%%%%
\begin{abstract}
In recent years, research about monitoring vital signs by smartphones grows significantly. There are some special sensors like Electrocardiogram (ECG) and Photoplethysmographic (PPG) to detect heart rate (HR) and respiration rate (RR). Smartphone cameras also can measure HR by detecting and processing imaging Photoplethysmographic (iPPG) signals from the video of a user’s face. Indeed, the variation in the intensity of the green channel can be measured by the iPPG signals of the video. This study aimed to provide a method to extract heart rate and respiration rate using the video of individuals' faces. The proposed method is based on measuring fluctuations in the  Hue, and can therefore extract both HR and RR from the video of a user's face. The proposed method is evaluated by performing on 25 healthy individuals. For each subject, 20 seconds video of his/her face is recorded. Results show that the proposed approach of measuring iPPG using Hue gives more accurate rates than the Green channel.
\end{abstract}


\section{Introduction}\label{sec:intro}
\begin{figure}[htbp] 
  \centering
  \includegraphics[width=1\columnwidth,height=2in]{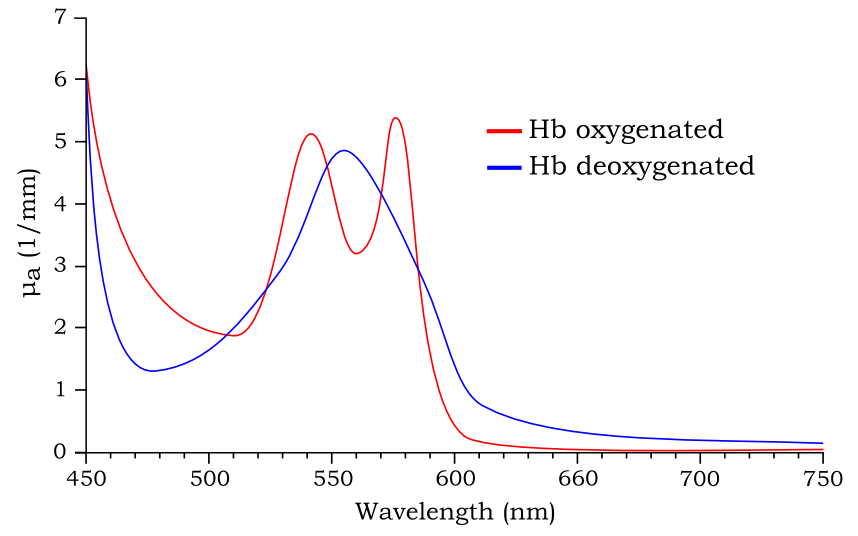}
  %\captionsetup{justification=centering}
  \caption{Absorbance Spectra of hemoglobin Hb (red) and oxygenated hemoglobin HbO2 (blue).}
  \label{fig:hb_wavelength}
  \end{figure}

Health tracking is one of the hot topics in the research areas. Today, there are many devices to detect biosignals. Thermometers, pulse monitors, pedometers, sleep trackers, calorie trackers, vein detectors, and blood sugar monitors among many other devices can be used to detect and monitor biosignals. Contact-based optical sensors are typically used to measure heart rate (HR). These sensors detect Photoplethysmographic (PPG) signal which is the variation of light reflectivity of different parts of the body (e.g., fingertip, wrist, ear, forehead) as a function of arterial pulsation. Different signal post-processing approaches \cite{zijlstra1991absorption, electrophysiology1996heart, tamura2014wearable, roald2013estimation, rolfe2000vivo} can extract HR and respiration rate (RR) from PPG signal. Indeed, the basis of these approaches is the fact that hemoglobin in the blood absorbs certain frequencies of light that can not be absorbed by surrounding tissues such as flesh and bone. As Fig. \ref{fig:hb_wavelength} shows, the desired wavelength varies from near-infrared (NIR) \cite{rolfe2000vivo}, to red \cite{roald2013estimation}. Due to the color variation in the skin, it is possible to extract physiological parameters like HR and RR using non-contact-based approaches. 

For the first time, Takano et al. showed that HR and RR can be extracted using a camera \cite{takano2007heart}. They calculated the average image brightness of the region of interest (ROI) in the images of the subject's skin. Applying some filters and performing spectral analysis, they extracted the HR, RR, and HRV (Heart Rate Variability). Kenneth et al. \cite{humphreys2007noncontact}, later improved the camera-based non-contact methods by capturing two iPPG signals simultaneously at two different wavelengths. In the same year, several authors used smartphone camera to extract non-contact physiological parameters like HR, HRV, and RR from facial video recorded in ambient light using three RGB channels \cite{mzahm2013agents, zhao2013remote, jimenez2013extracting, gubbi2013internet, foteinos2013cognitive, datcu2013noncontact, parnandi2013contactless, shao2015noncontact, lewandowska2011measuring, li2014remote}. Another non-contact-based physiological parameter extraction system has been proposed by Rahman et al. \cite{rahman2015non}. They used a simple laptop web camera to detect HR, RR, and IBI (inter-bit interval) with about 90\% accuracy. From most of the related literature, most of the non-contact systems to monitor physiological parameters are done offline, and most of them are good for a certain amount of time in the lab environment. However, in \cite{rahman2016real}, authors presented a non-contact HR monitoring system in real-time for an unlimited amount of time using a web camera. To overcome the effect of ambient noise in the green channel, Sanyal et al. \cite{sanyal2018algorithms} proposed a novel Hue (HSV color space) based observable for reflection-based iPPG.

This paper presents a non-contact method to extract HR and RR in real-time based on video HSV analysis. By measuring the variations of the average Hue, arterial pulsations can be extracted from the user's forehead. To this end, we first detect the user’s face and eyes, then detect his/her forehead based on the face and eyes location. Then we find the variation in the Hue of the same areas in different frames and convert the data to the frequency domain. Since the frequency of HR and RR are in a specific range, we can calculate the HR and RR.

The rest of this paper is organized as follows. In Section II, a detailed description of the steps that are needed to extract the HR and RR from a real-time video. We then show the result of our experiment and demonstrate the accuracy of our method by comparing the results with actual HR and RR rates in Section III. Finally, the conclusions and future work are drawn in Section IV.

\section{System Design} \label{sec:sysmodel}
In a contract-based approach, the PPG sensor illuminates the subject's skin with a LED (usually green color). It detects the absorption of light due to the arterial pulsation by a photodiode [15].
However, in non-contact-based methods, a camera is used to track a specific RGB channel (e.g., the Green channel), at which oxygenated hemoglobin absorbs some particular frequencies of the light that can not be absorbed by the surrounding tissue. These methods measure the iPPG signal, which is usually the average fluctuation of the green channel of all the frames in the video. Since noise can affect the fluctuation of the green channel, we replaced the RGB channel with the HSV color space convert and considered the Hue of all the pixels in all video frames. Our method's phases are as follows.

\subsection{Forehead Detection} 
\begin{figure} [htbp]
  \centering
  \includegraphics[width=1\columnwidth,height=2.8in]{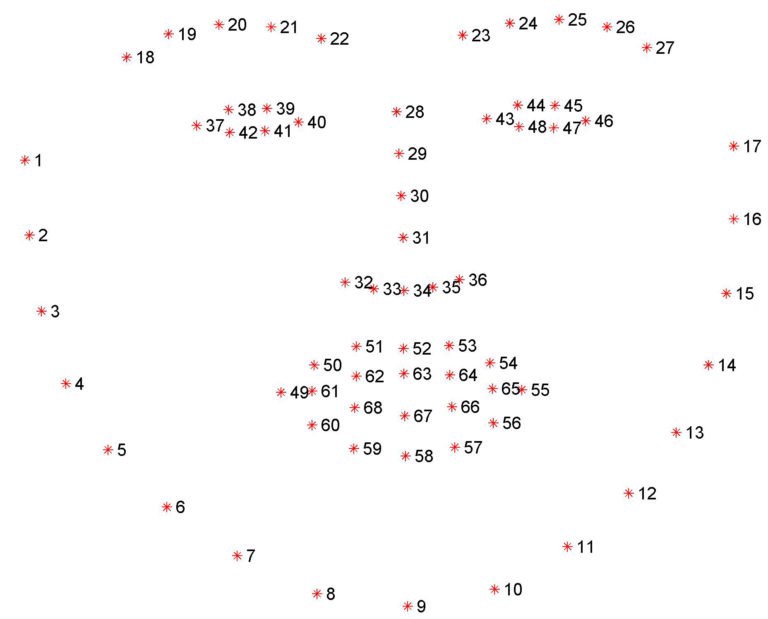}
  \caption{Dlib's 68 points (landkmarks) of the face.} \label{fig: facial_landmarks}
\end{figure}

Color variation due to the arterial pulsation can be detected from the forehead area. Hence, the first task is to detect the forehead in the video frame. To this end, we can detect the face and 
eye/eyebrow, then calculate the position of the forehead based on them. We use Dlib library to detect face and facial landmarks. Dlib exploits a face detection model based on Histogram of Oriented Gradients (HoG) features and Support Vector Machine (SVM) \cite{dalal2005histograms}. This method extracts features (HoG) into a vector and feeds them into an SVM classification algorithm to detect faces (or other trained objects) in an image. After detecting the face, we used Dlib to detect facial landmarks that estimate the location of 68 coordinates $(x, y)$ that map the facial points on a person’s face (Fig. \ref{fig: facial_landmarks}). We then considered the area between face's rectangular and points $(21, 24)$ of the facial landmarks as forehead.

\begin{figure} [htbp]
  \centering
  \includegraphics[width = 11pc]{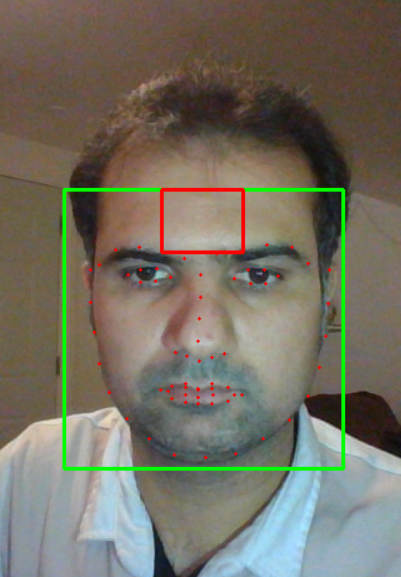}
  \caption{Foreheade detection based on face detection and facial landmarks $(21, 24)$} \label{fig: face_rec}
\end{figure}

\subsection{Average Hue Calculation}
In this phase, we need to calculate the average Hue of pixels in each frame. In our experiments, we use a laptop webcam that can capture 30 RGB frames per second. Since our purpose is to calculate HR and RR in real-time, we have to pay its cost, losing some frames due to the processing time. Using Dlib library and a common laptop CPU, finally, we have around nine frames per second. Hence, we need to convert the RGB frame to the HSV frame and calculate the average Hue for each frame. Sanyal et al. in \cite{sanyal2018algorithms} demonstrated that considering Hue within the range of (0, 0.1), it is possible to measure the variations corresponding to the skin color. Therefore, to consider only pixels with Hue in (0, 0.1), we ignored other pixels. There are nine average values per second from all the frames, which demonstrate the change in the Hue of the forehead's pixels that have Hue from 0 to 0.1 corresponding to the color variation due to the arterial pulsation. Indeed, these average values construct the iPPG signal.

\begin{figure}[htbp] 
  \centering
  \includegraphics[width = 11pc]{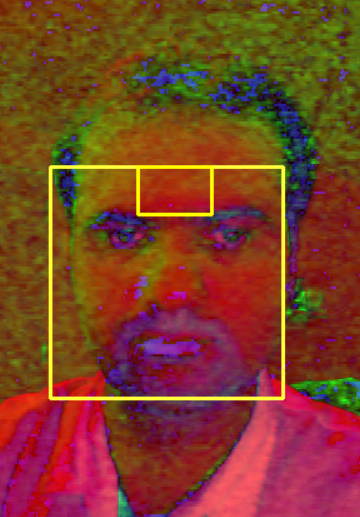}
  \caption{HSV color space of detected face and forehead.} \label{fig: hsv}
\end{figure}

\subsection{Spectrum analysis}
\begin{figure} [htbp] 
  \centering
\subfloat[\label{iPPG-t-f_a}]{%
     \includegraphics[width=0.9\linewidth]{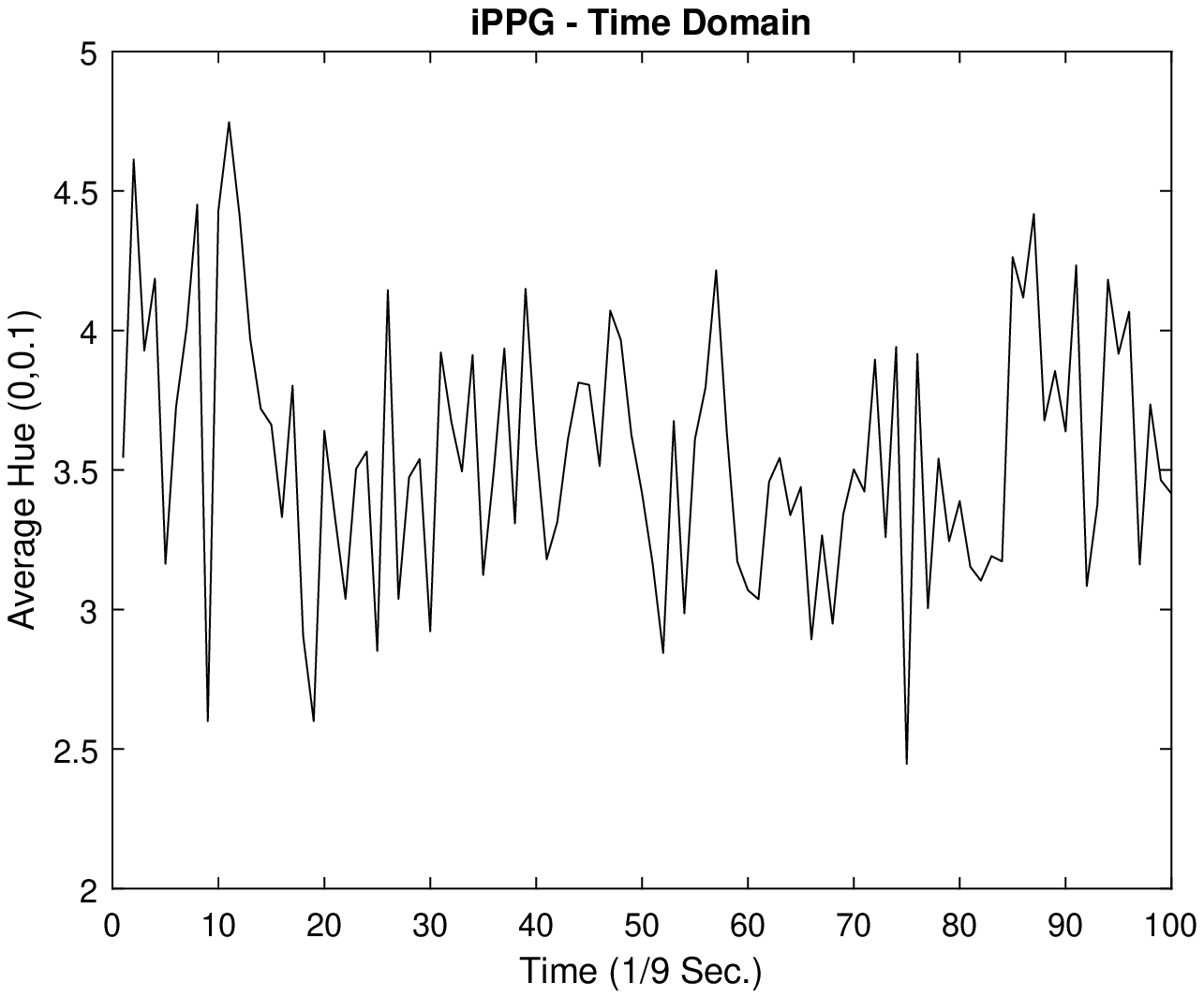}}
  \hfill
% \subfloat[\label{1b}]{%
%       \includegraphics[width=0.45\linewidth]{HR_time.eps}}
  \\
\subfloat[\label{iPPG-t-f_b}]{%
      \includegraphics[width=0.9\linewidth]{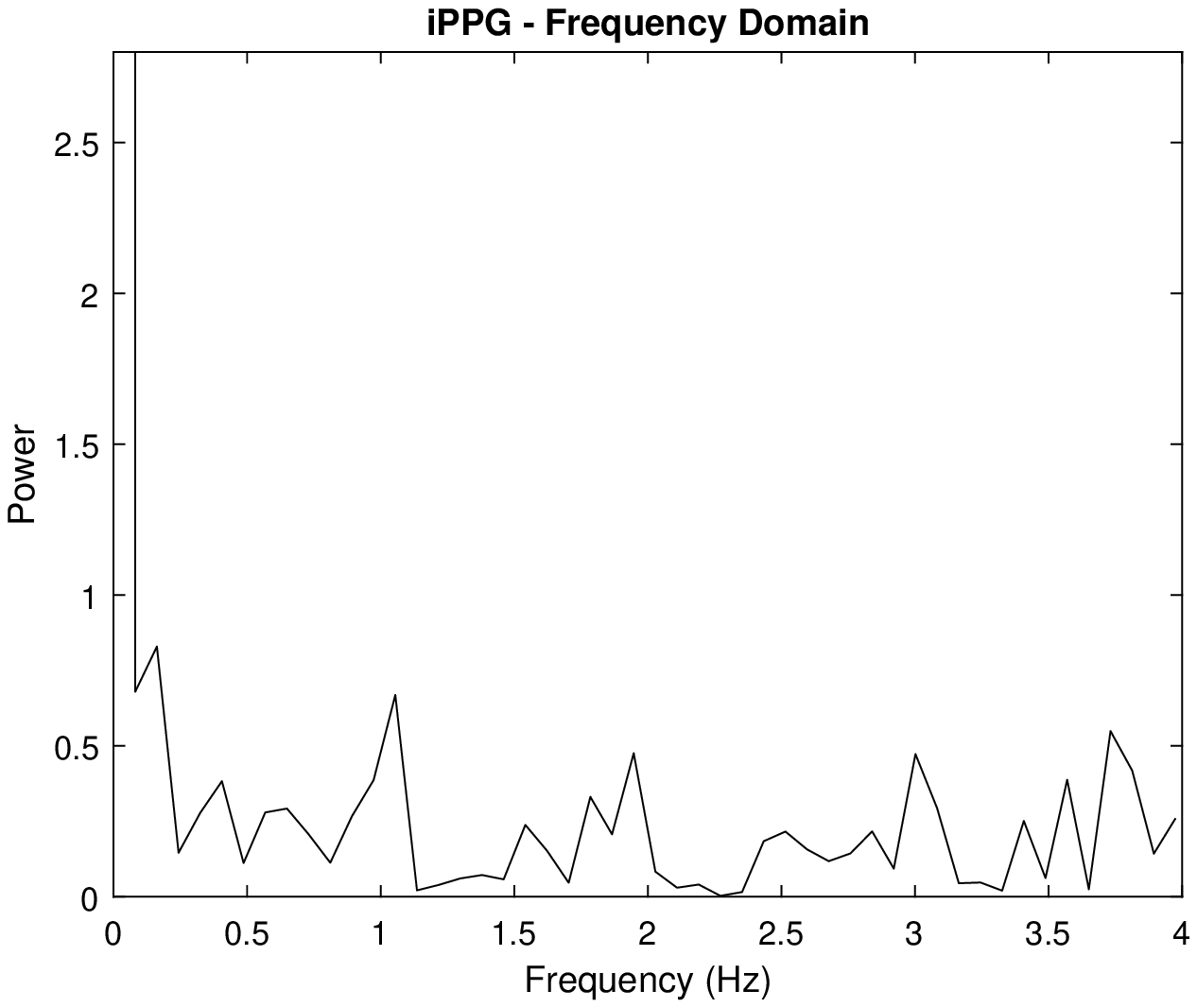}}
  \hfill
% \subfloat[\label{1d}]{%
%       \includegraphics[width=0.45\linewidth]{HR_time.eps}}

\caption{Raw iPPG signal obtained from a laptop webcam in 10 seconds (a) time domain, (b) frequency domain.}
\label{iPPG-t-f} 
\end{figure}

To detect the HR and RR from iPPG signal, we need to convert this signal from the time domain to the frequency domain and extract the frequencies corresponding to HR, and RR. Fig. \ref{iPPG-t-f} shows both the time and frequency domain of a raw iPPG signal obtained from an 11-second recording of a user's face by a regular laptop webcam which has resolution 640*480 pixels and captures 30 frames per second. Although we have 30 frames per second, 21 frames will be drop due to the real-time processing because each frame will be processed in real-time. Therefore, 99 frames will be processed in 11 seconds.

\begin{figure} [htbp]
  \centering
\subfloat[\label{HR-RR_signals_a}]{%
     \includegraphics[width=0.43\linewidth]{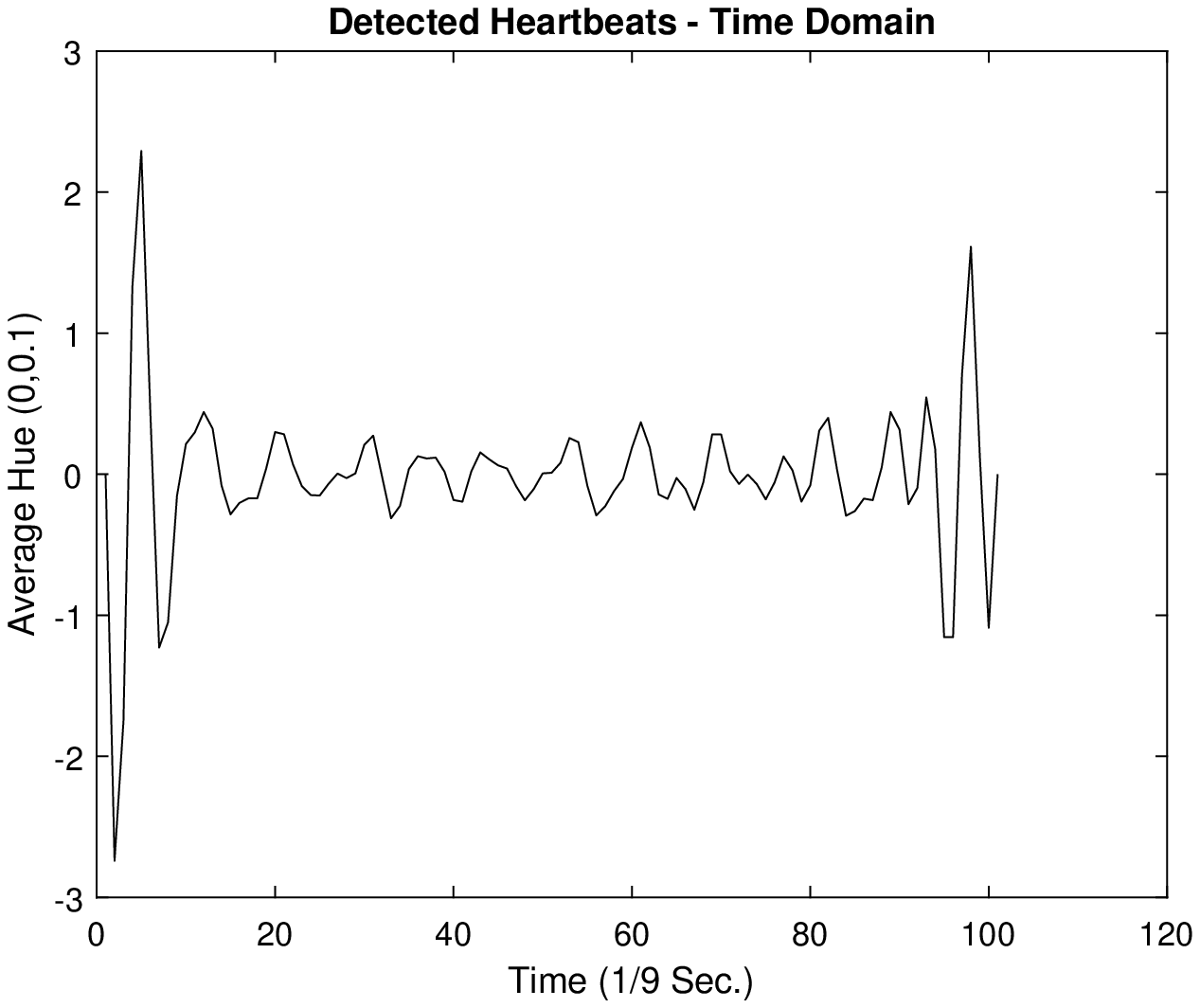}}
  \hfill
\subfloat[\label{HR-RR_signals_b}]{%
      \includegraphics[width=0.43\linewidth]{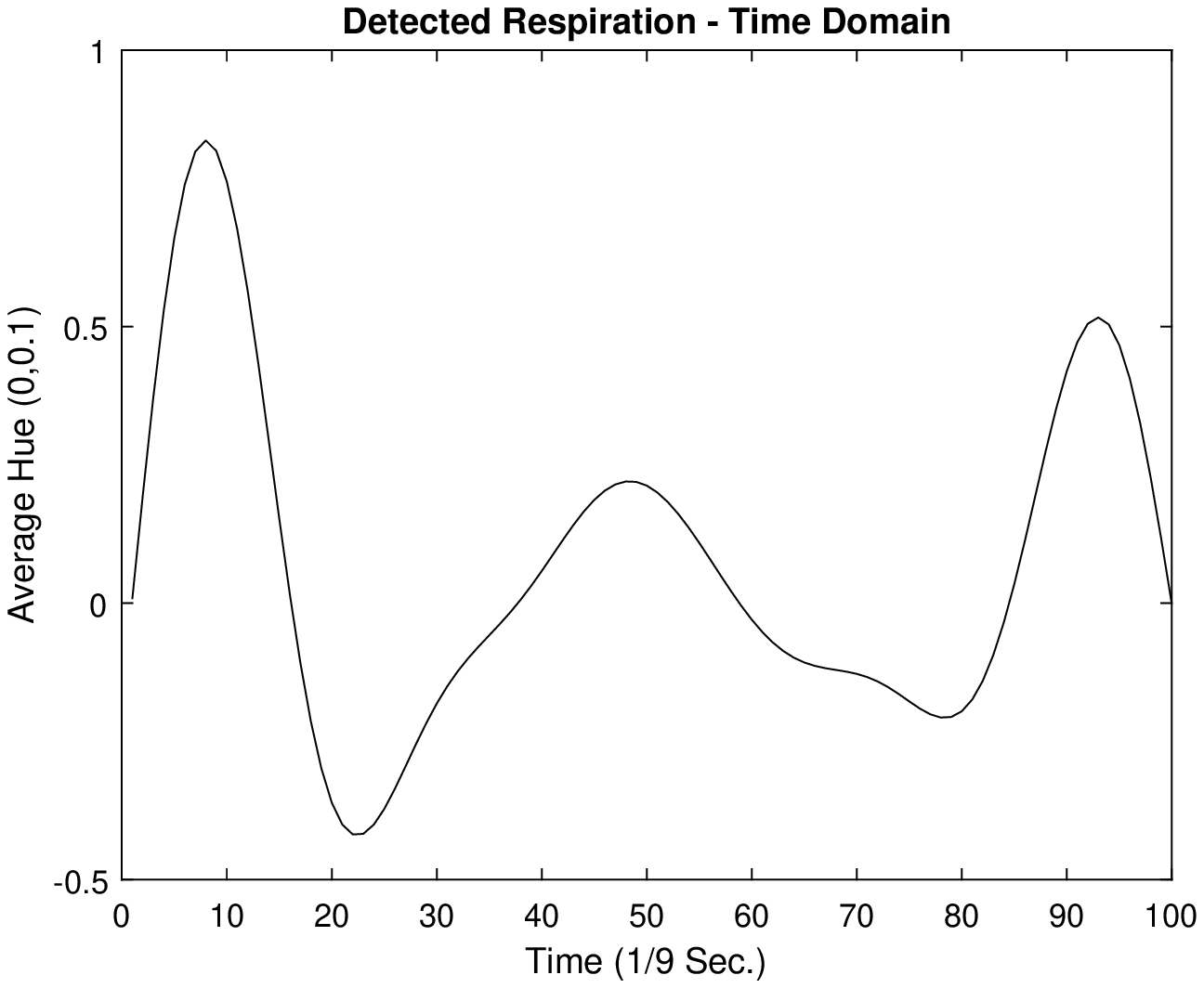}}
  \\
\subfloat[\label{HR-RR_signals_c}]{%
      \includegraphics[width=0.43\linewidth]{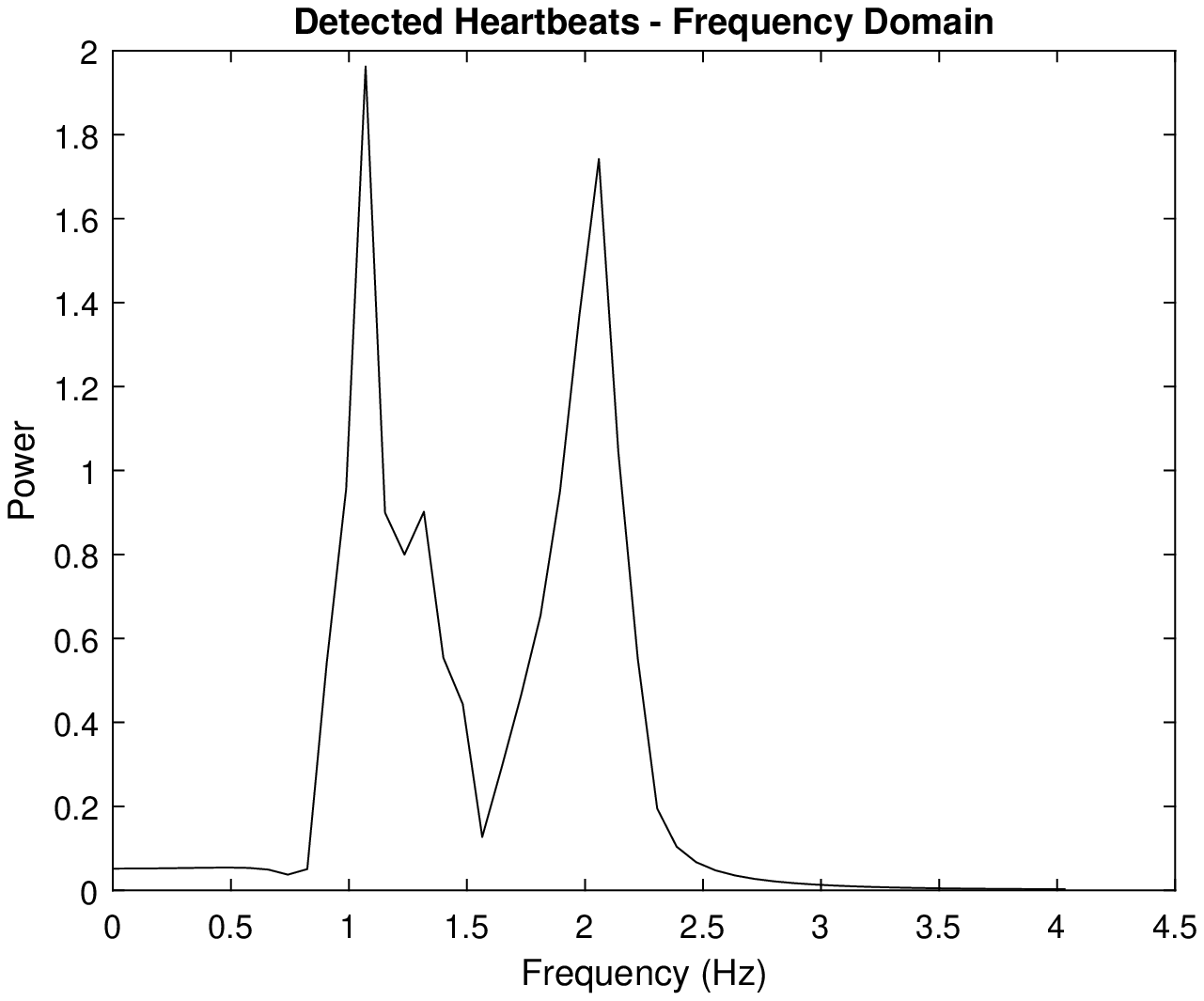}}
  \hfill
\subfloat[\label{HR-RR_signals_d}]{%
      \includegraphics[width=0.43\linewidth]{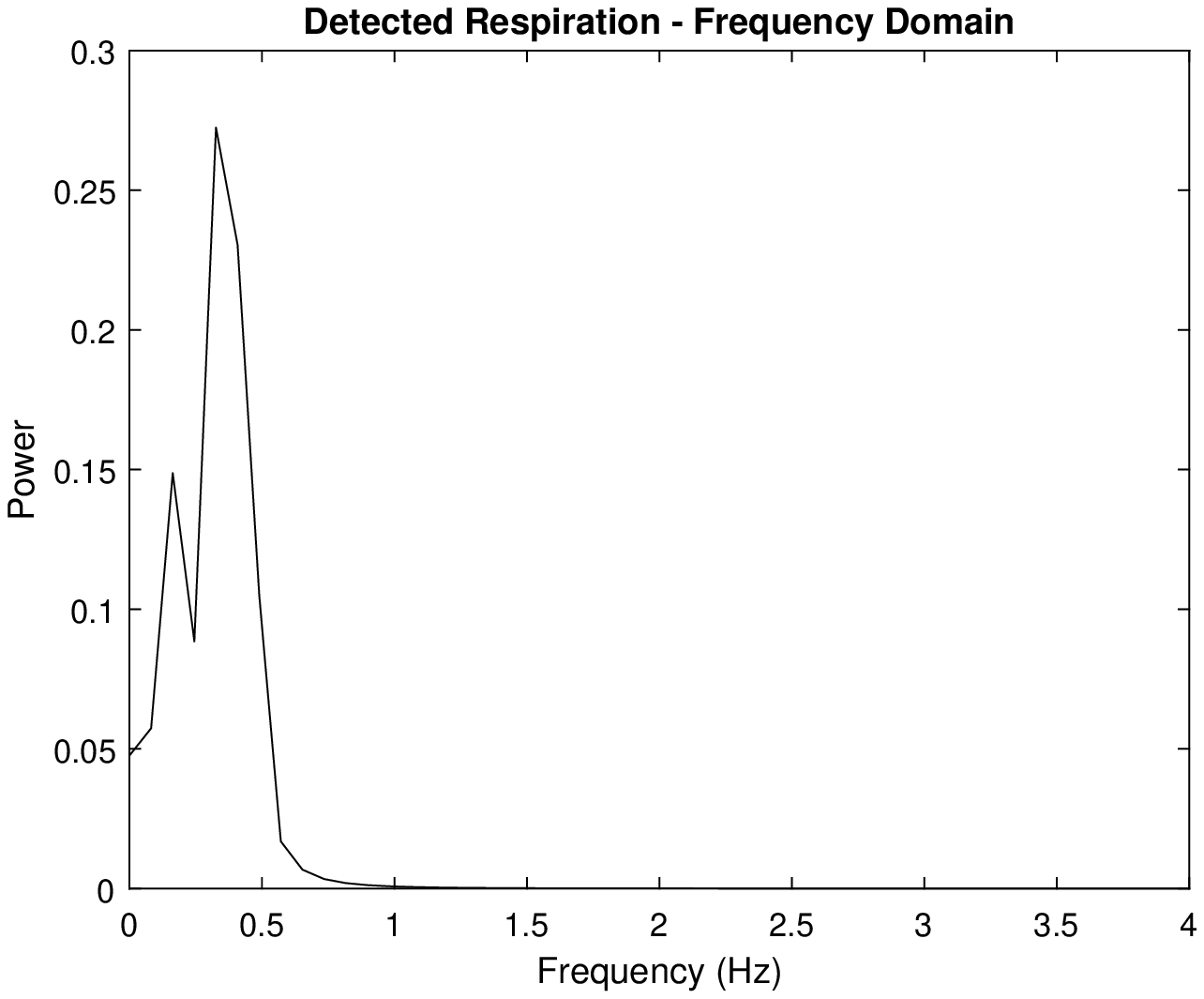}}

\caption{Heartbeat and respiration signal extracted from iPPG signal using band-pass filters (0.8, 2) and (0.18, 0.5) (a) Heartbeat - time domain, (b) Respiration - time domain (c) Heartbeat - frequency domain (d) Respiration - frequency domain.}
\label{HR-RR_signals} 
\end{figure}

\begin{figure}
  \centering
  \includegraphics[width=1\linewidth]{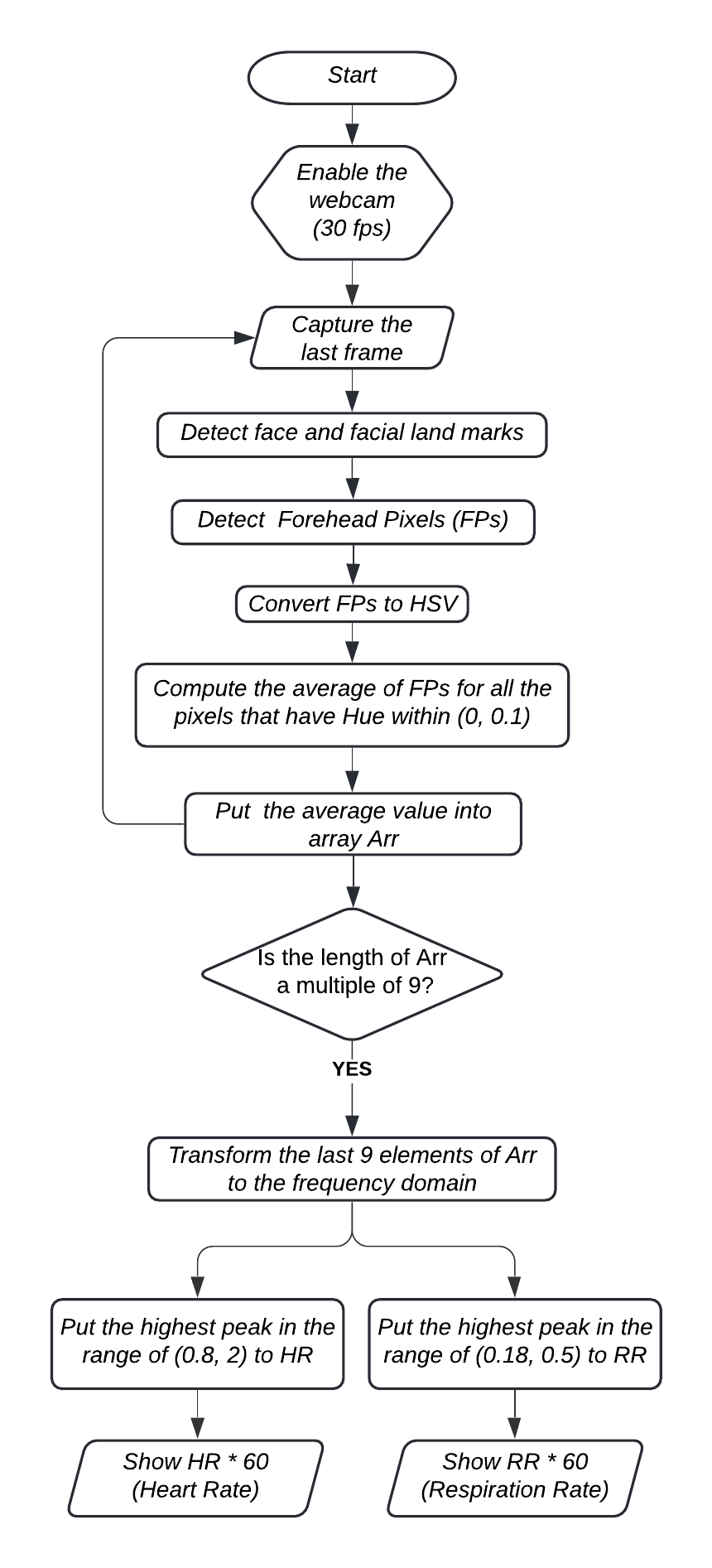}
  \caption{flowchart of the algorithm for extracting HR and RR from facial video.} \label{fig: flowchart}
\end{figure}

The frequency of HR and RR typically are in the range of (0.8, 2.2) and (0.18, 0.5), respectively \cite{sanyal2018algorithms}. Therefore, it is possible to separate these signals from the time domain iPPG signal using band-pass filters with cutoff frequency in the range of HR and RR frequencies. Fig. \ref{HR-RR_signals} shows the iPPG signal in both frequency and time domain after passing throughout band-pass filters with cutoff frequencies (0.8, 2.2) and (0.18, 0.5). Finally, the peak of filtered frequency Spectra in the mentioned ranges represents HR and RR. As Fig. \ref{HR-RR_signals_c} shows, the maximum peak in the range of (0.8, 2.2) is 1.1 Hz, so the HR is 66. The corresponding heartbeat signal is shown in Fig. \ref{HR-RR_signals_a}. Also, we can infer that RR is 18 because the maximum peak in the range of (0.18, 0.5) is 3.3 Hz, as shown in Fig. \ref{HR-RR_signals_d}. Fig. \ref{HR-RR_signals_b} shows the corresponding breathing signal. A flowchart of our algorithm is provided in Fig. \ref{fig: flowchart}.

\section{Experiment results}
\begin{figure}
  \centering
  \includegraphics[width = 20.5pc]{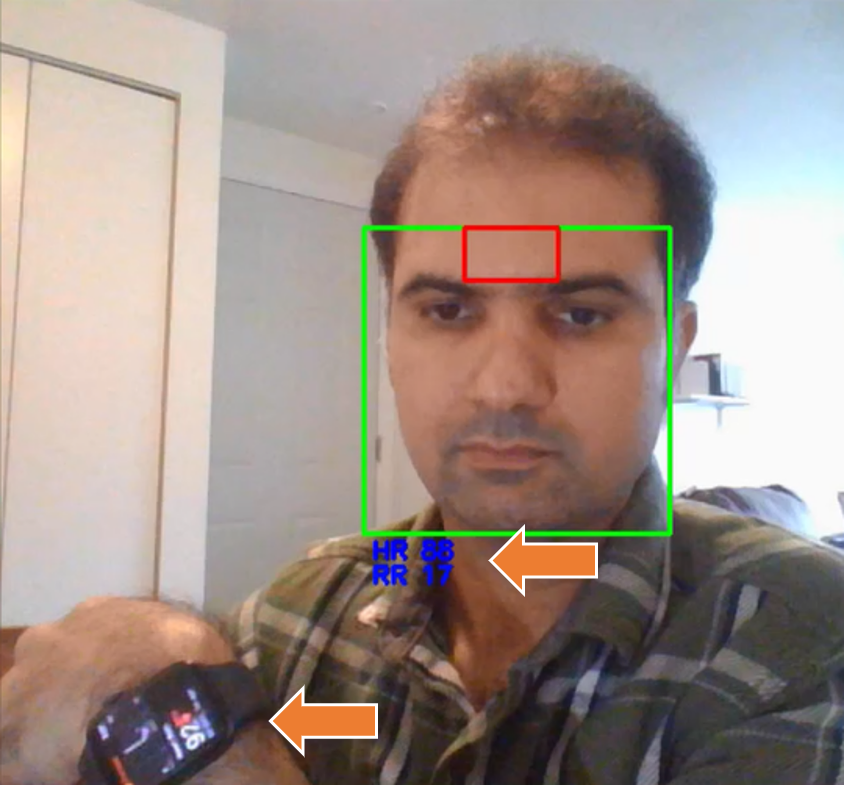}
  \caption{Real time iPPG-based detected HR is 88 while Apple watch 6 Heart Rate app shows 92 as HR value.} \label{fig: compare_apple}
\end{figure}

The computing platform used is an HP Pavilion laptop model 15-eg0, which has a webcam with a resolution of 640*480 pixels, and it is capable of capturing 30 frames per second. All implementations are performed on Python version 3.7.7 that initially takes about 2 seconds to estimate the HR and 6 seconds to detect the RR after running the codes. It needs this initial delay because each heartbeat takes between 0.45 to 1.25 seconds and each consecutive inhalation and exhalation takes between 2 to 5.56 seconds. After the initial estimation, we considered the average of all detected HR and RR within the last second. To show more stable values on the screen, we take an average over ten last HR and RR. To compare the result with other contact-based devices, we used two smartwatches to detect HR using PPG sensors. Accuracy. A Samsung Watch 3 and an Apple watch 6 are used for HR monitoring (Fig. \ref{fig: compare_apple}). We also used a Hexoskin smart shirt to track the actual RR and use its ECG signal to have accurate HR as the primary reference. Hexoskin uses the respiratory inductance plethysmography (RIP) method to evaluate pulmonary ventilation by measuring the chest and abdominal wall movement. Hence, it can provide an accurate RR as a reference signal. 
On the other hand, to compare the performance of our methods with previous iPPG methods, we also calculate the HR and RR based on the average of the green channel from each frame. We collected data of all devices for 100 seconds and compared HR and RR every 5 seconds which can be seen in Table \ref{tbl_comparison}.

\begin{table}
  \centering
  \caption{Detected HR and RR from non-contact-based and contact-based methods.}
  \begin{tabular}{|p{.5cm}|p{.6cm}|p{.6cm}|p{.9cm}|p{.3cm}|p{.6cm}||p{.6cm}|p{.3cm}|p{.55cm}|}
    \hline
   & \multicolumn{4}{c}{\textbf{HR}} & \textbf{} & \multicolumn{2}{c}{\textbf{RR}} & \textbf{}\\
   \hline
  \textbf{Time (sec.)} & \textbf{Hexo- skin} & \textbf{Apple watch} & \textbf{Samsung watch} & \textbf{Hue} & \textbf{Green} & \textbf{Hexo- skin} & \textbf{Hue} & \textbf{Green} \\
  \hline
  10 & \textbf{72} & 74 & 72 & 66 & 70 & \textbf{18} & 16 & 16 \\
15 & \textbf{70} & 73 & 71 & 68 & 72 & \textbf{18} & 19 & 17 \\
20 & \textbf{71} & 75 & 73 & 69 & 72 & \textbf{16} & 18 & 17 \\
25 & \textbf{70} & 75 & 72 & 71 & 65 & \textbf{17} & 18 & 15 \\
30 & \textbf{69} & 74 & 71 & 71 & 69 & \textbf{15} & 19 & 19 \\
35 & \textbf{69} & 73 & 71 & 70 & 72 & \textbf{17} & 18 & 18 \\
40 & \textbf{70} & 73 & 71 & 69 & 75 & \textbf{16} & 18 & 18 \\
45 & \textbf{70} & 74 & 72 & 66 & 74 & \textbf{18} & 18 & 16 \\
50 & \textbf{69} & 74 & 73 & 68 & 70 & \textbf{19} & 17 & 13 \\
55 & \textbf{71} & 73 & 71 & 72 & 68 & \textbf{17} & 17 & 16 \\
60 & \textbf{70} & 73 & 70 & 73 & 73 & \textbf{17} & 18 & 17 \\
65 & \textbf{72} & 75 & 71 & 76 & 69 & \textbf{18} & 19 & 16 \\
70 & \textbf{73} & 74 & 71 & 75 & 68 & \textbf{16} & 18 & 18 \\
75 & \textbf{73} & 72 & 70 & 70 & 72 & \textbf{17} & 18 & 18 \\
80 & \textbf{75} & 72 & 74 & 69 & 70 & \textbf{17} & 20 & 22 \\
85 & \textbf{74} & 76 & 73 & 71 & 75 & \textbf{18} & 18 & 20 \\
90 & \textbf{74} & 75 & 71 & 72 & 74 & \textbf{16} & 18 & 18 \\
95 & \textbf{75} & 75 & 73 & 72 & 76 & \textbf{17} & 17 & 19 \\
100 & \textbf{76} & 75 & 72 & 73 & 75 & \textbf{17} & 17 & 18 \\
  \hline
  \end{tabular}
  \label{tbl_comparison}
  \end{table}

Considering the root mean square error (RMSE) as a performance metric, Table~\ref{tbl_RMSE} provides a quantitative comparison of different HR detection methods. From the results, it can be seen that the detected HR by average Hue approach has 2.95 RMSE average, which is less than 5, the standard RMSE for HR monitors as set by Advancement of Medical Instrumentation EC-13. Also, comparing the results of the Hue and green channel reveals that the proposed method has provide promising accuracy among the other competitors.
  % Please add the following required packages to your document preamble:
% \usepackage{multirow}
%\vspace{-.1cm}
\begin{table}[H]
  \centering
  \caption{Average RMES of different devices in comparison with refrence value of Hexoskin.}
  \begin{tabular}{|p{.8cm}|p{.7cm}|p{1cm}|l|l||l|l|}
    \hline
  \multirow{2}{*}{} & \multicolumn{3}{c}{\textbf{HR}} & \textbf{}& \multicolumn{1}{c}{\textbf{RR}} & \textbf{} \\
  \hline
   & \textbf{Apple watch} & \textbf{Samsung   watch} & \textbf{Hue} & \textbf{Green} & \textbf{Hue} & \textbf{Green} \\
   \hline
  \textbf{Average RMSE} & 3.1119 & 2.0901 & 2.9558 & 3.0262 & 1.7014 & 2.5026 \\
  \hline
  \end{tabular}
  \label{tbl_RMSE}
  \end{table}

\section{Conclusion}\label{conclude}
Monitoring the heart rate and respiration rate by non-contact methods usually rely on the fluctuation of a particular RGB color space, like the green channel. While this paper has provided a real-time HR and RR monitoring method based on the change in the Hue channel in the HSV color space. The experiments were performed on the user's facial video in real-time to extract the HR and RR. The evaluation results as well as comparison with other HR detection algorithms have shown that our approach has a lower RMSE (2.95 for HR and 1.70 for RR), and have proven that the effectiveness of our method for real-time HR and RR monitoring. In addition, the computational efficiency indicating the potential of the proposed method for the real-time applications. The research work is progressing to investigate the impact of different colors of the skin on the ability of the proposed technique.

\bibliographystyle{IEEE}
\bibliography{keylatex}
\end{document}